\documentclass[aps,preprint,preprintnumbers,amsmath,amssymb,nofootinbib]{revtex4}
\usepackage{amssymb}
\usepackage{lscape}
\usepackage{amsmath}
\usepackage{bm}
\usepackage{graphicx}
\usepackage{epsfig}
\begin{document}
\title{Inflationary back-reaction effects from Relativistic Quantum Geometry}
\author{$^{1,2}$ Mauricio Bellini\footnote{E-mail address: mbellini@mdp.edu.ar} }
\address{$^1$ Departamento de F\'isica, Facultad de Ciencias Exactas y
Naturales, Universidad Nacional de Mar del Plata, Funes 3350, C.P.
7600, Mar del Plata, Argentina.\\
$^2$ Instituto de Investigaciones F\'{\i}sicas de Mar del Plata (IFIMAR), \\
Consejo Nacional de Investigaciones Cient\'ificas y T\'ecnicas
(CONICET), Mar del Plata, Argentina.}

\begin{abstract}
We study the dynamics of scalar metric fluctuations in a non-perturbative variational formalism recently introduced, by which
the dynamics of an geometrical scalar field $\theta$, describes the quantum geometrical effects on a Weylian-like manifold with respect to a background Riemannian space-time.
In this letter we have examined an example in the framework of inflationary cosmology. The resulting spectral predictions are in very good agreement with observations and other models of inflation.
\end{abstract}
\maketitle

\section{Introduction}

The inflationary theory of the universe provides a physical mechanism to
generate primordial energy density fluctuations on cosmological scales \cite{infl}. The primordial
scalar perturbations drive the seeds of large scale structure which then had gradually formed today's galaxies, which is being
tested in current observations of cosmic microwave background (CMB). These fluctuations are today larger than a thousand size of a typical galaxy, but during inflation
were very much larger than the size of the causal horizon. According with this scenario, the almost constant potential depending of a minimally coupled to
gravity inflation field $\varphi$, called the inflaton, caused the accelerated expansion of the very early universe.
In particular, back-reaction effects has been subject of study. Quantum vacuum fluctuations are continuously generated on sub-Hubble scales. As the wavelengths of these fluctuation modes exit the Hubble radius, the vacuum oscillations the modes get squeezed and become the seeds for the observed inhomogeneities in the distribution of matter and anisotropies. In this framework, the evolution of scalar metric fluctuations has been studied in\cite{ab}.

In this letter, I consider gauge-invariant fluctuations of the metric using a new variational method recently introduced named Relativistic Quantum Geometry (RQG).
These fluctuations were extensively studied using a linear perturbative corrections\cite{mio}. Nonlinear perturbative corrections were studied in \cite{muk}.
The scalar metric perturbations of the metric are associated with the density perturbations. These are spin-zero projections of the graviton, which only exist in non-vacuum cosmology. The issue of gauge-invariance becomes critical when we attempt to analyse how the scalar metric perturbations produced in the very early universe influences the global flat, isotropic and
homogeneous universe, described by a background FLRW metric.

\section{Geometrical quantum dynamics}

The variation of the
metric tensor must be done in a Weylian-like
integrable manifold\cite{RB} using an
auxiliary geometrical scalar field $\theta$, in order to the
Einstein tensor (and the Einstein equations) can be represented on
a Weylian-like manifold\cite{Weyl}, in agreement with the gauge-invariant
transformations obtained in \cite{RB}. If we consider a zero covariant derivative of the metric tensor in the Riemannian manifold
(we denote with a semicolon the Riemannian-covariant derivative): $\Delta g_{\alpha\beta}=g_{\alpha\beta;\gamma} \,dx^{\gamma}=0$,
hence the Weylian-like covariant derivative $ g_{\alpha\beta|\gamma} = \theta_{\gamma}\,g_{\alpha\beta}$, described with respect to the Weylian-like connections \footnote{To simplify the notation we shall denote $\theta_{\alpha} \equiv \theta_{,\alpha}$}
\begin{equation}\label{gama}
\Gamma^{\alpha}_{\beta\gamma} = \left\{ \begin{array}{cc}  \alpha \, \\ \beta \, \gamma  \end{array} \right\}+ \theta^{\alpha} \hat{g}_{\beta\gamma} ,
\end{equation}
will be nonzero
\begin{equation}\label{gab}
\delta \hat{g}_{\alpha\beta} = \hat{g}_{\alpha\beta|\gamma} \,dx^{\gamma} = -\left[\theta_{\beta} \hat{g}_{\alpha\gamma} +\theta_{\alpha} \hat{g}_{\beta\gamma}
\right]\,dx^{\gamma}.
\end{equation}
From the action's point of view, the scalar field $\theta(x^{\alpha})$ is a generic geometrical transformation that leads invariant the action\cite{rb}
\begin{equation}\label{aac}
{\cal I} = \int d^4 \hat{x}\, \sqrt{-\hat{g}}\, \left[\frac{\hat{R}}{2\kappa} + \hat{{\cal L}}\right] = \int d^4 \hat{x}\, \left[\sqrt{-\hat{g}} e^{-2\theta}\right]\,
\left\{\left[\frac{\hat{R}}{2\kappa} + \hat{{\cal L}}\right]\,e^{2\theta}\right\},
\end{equation}
where we denote with a {\em hat}, the quantities represented on the semi-Riemannian manifold. Hence, Weylian-like quantities will be varied over these
quantities in a semi-Riemannian manifold so that the dynamics of the system preserves the action: $\delta {\cal I} =0$, and we obtain
\begin{equation}
-\frac{\delta V}{V} = \frac{\delta \left[\frac{\hat{R}}{2\kappa} + \hat{{\cal L}}\right]}{\left[\frac{\hat{R}}{2\kappa} + \hat{{\cal L}}\right]}
= 2 \,\delta\theta,
\end{equation}
where $\delta\theta = \theta_{\mu} dx^{\mu}$ is an exact differential and $V=\sqrt{-\hat{ g}}$ is the volume of the Riemannian manifold. Of course, all the variations are in the Weylian-like geometrical representation, and assure us gauge invariance because $\delta {\cal I} =0$.
Using the fact that the tetra-length is given by $S=\frac{1}{2} x_{\nu} \hat U^{\nu}$ and the Weylian-like velocities are given by $u^{\mu} = \hat U^{\mu} + 2\theta^{\mu} S$, can be demonstrated that
\begin{equation}
u^{\mu} u_{\mu} = 1 + 4 S \left( \theta_{\mu} \hat U^{\mu} - \frac{4}{3} \Lambda\, S\right).
\end{equation}
The components $u^{\mu}$ are the relativistic quantum velocities, given by the geodesic equations
\begin{equation}
\frac{du^{\mu}}{dS} + \Gamma^{\mu}_{\alpha\beta} u^{\alpha} u^{\beta} =0,
\end{equation}
such that the Weylian-like connections $\Gamma^{\mu}_{\alpha\beta}$ are described by (\ref{gama}). In other words, the quantum velocities
$u^{\mu}$ are transported with parallelism on the Weylian-like manifold, meanwhile $\hat{U}^{\mu}$ are transported with parallelism on the Riemann manifold.
If we require that $u^{\mu} u_{\mu} = 1$, we obtain the gauge
\begin{equation}\label{gau}
\hat\nabla_{\mu} A^{\mu} =-2 \frac{d\theta}{dS}.
\end{equation}
Since was demonstrated in \cite{RB}
the Einstein tensor can be written as
\begin{equation}
\bar{G}_{\alpha\beta} = \hat{G}_{\mu\nu} + \theta_{\alpha ; \beta} + \theta_{\alpha} \theta_{\beta} + \frac{1}{2} \,g_{\alpha\beta}
\left[ \left(\theta^{\mu}\right)_{;\mu} + \theta_{\mu} \theta^{\mu} \right],
\end{equation}
and we can obtain the invariant cosmological constant $\Lambda$
\begin{equation}\label{p}
\Lambda = -\frac{3}{4} \left[ \theta_{\alpha} \theta^{\alpha} + \hat{\Box} \theta\right],
\end{equation}
so that we can define a geometrical Weylian-like quantum action
${\cal W} = \int d^4 x \, \sqrt{-\hat{g}} \, \Lambda$, such that the dynamics of the geometrical field, after imposing $\delta
W=0$, is described by the Euler-Lagrange equations which take the form
\begin{equation}\label{q}
\hat{\nabla}_{\alpha} \Pi^{\alpha} =0, \qquad {\rm or} \qquad \hat\Box\theta=0,
\end{equation}
where the momentum components are $\Pi^{\alpha}\equiv -{3\over 4} \theta^{\alpha}$ and the relativistic quantum algebra is given by\cite{RB}
\begin{equation}\label{con}
\left[\theta(x),\theta^{\alpha}(y) \right] =- i \Theta^{\alpha}\, \delta^{(4)} (x-y), \qquad \left[\theta(x),\theta_{\alpha}(y) \right] =
i \Theta_{\alpha}\, \delta^{(4)} (x-y),
\end{equation}
with $\Theta^{\alpha} = i \hbar\, \hat{U}^{\alpha}$ and $\Theta^2 = \Theta_{\alpha}
\Theta^{\alpha} = \hbar^2 \hat{U}_{\alpha}\, \hat{U}^{\alpha}$ for the Riemannian components of velocities $\hat{U}^{\alpha}$.

\section{Power-law inflation}

In order to describe an example, we shall consider the case of an inflationary universe where the scale factor of the universe describes a power-law expansion, and the line
element related with the background semi-Riemannian curvature, is
\begin{equation}
d\hat{S}^2 = \hat{g}_{\mu\nu} d\hat{x}^{\mu} d\hat{x}^{\nu}= d\hat{t}^2 - a^2(t) \hat{\eta}_{ij} d\hat{x}^i d\hat{x}^j,
\end{equation}
where the {\em hat} denotes that the metric tensor es defined over a semi-Riemannian manifold. We shall define the action ${\cal I}$
on this manifold, so that the background action describes the expansion driven by a scalar field, which is minimally coupled to gravity
\begin{equation}
{\cal I} = \int \, d^4x\, \sqrt{-\hat{g}}\, \left[ \frac{{\cal \hat{R}}}{16 \pi G} +  \left[ \frac{1}{2}\dot\phi^2 - V(\phi)\right]\right],
\end{equation}
In power-law inflation the scale factor of the universe and the Hubble parameter, are given respectively by\cite{prd96}
\begin{equation}
a(t)= \beta \, t^p, \qquad H(t)= \frac{p}{t},
\end{equation}
where $\beta= {a_0 \over t^p_0}$, $a_0$ is the initial value of the scale factor, $t_0$ is the initial value of the cosmic time, and the background solution for the
inflaton field dynamical equation
\begin{equation}
\ddot\phi + 3 \frac{\dot{a}}{a} \dot\phi + V'(\phi) =0,
\end{equation}
is
\begin{equation}
\phi(t) = \phi_0\left[1 - {\bf ln} \left(\frac{\alpha }{4 \pi \phi^2_0\,G} t\right)\right],
\end{equation}
where $p = 4\pi G \phi^2_0$, $\beta = \frac{a_0}{t^p_0}$ and $\alpha = H_f$ is the value of the Hubble parameter at the end of inflation.
The scalar potential can be written in terms of the scalar field
\begin{equation}
V(\phi) = \frac{3}{8\pi G H^2_f} \left(1 - \frac{1}{12 \pi G \phi^2_0}\right) e^{2 (\phi/\phi_0)},.
\end{equation}
which decreases with $\phi$.

\subsection{Geometrical dynamics of space-time}

The geometrical scalar field $\theta$ can be expressed as a Fourier expansion
\begin{equation}
\theta(\vec{x},t) = \frac{1}{(2\pi)^{3/2}} \int \, d^3k \, \left[ A_k \, e^{i \vec{k}.\vec{x}} \xi_k(t) + A^{\dagger} \, e^{-i \vec{k}.\vec{x}} \xi^*_k(t) \right],
\end{equation}
where $A^{\dagger}$ and $A_k$ are the creation and annihilation operators. From the point of view of the metric tensor, an example in power-law inflation can be illustrated by
\begin{equation}
g_{\mu\nu} = {\rm diag}\left[ e^{2\theta}, - a^2(t) e^{-2\theta}, - a^2(t) e^{-2\theta}, - a^2(t) e^{-2\theta}\right],
\end{equation}
such that the related quantum volume is $V_q= a^3(t) e^{-2\theta}= \sqrt{-\hat{g}} \,e^{-2\theta}$. The dynamics for $\theta$ is governed by the equation
\begin{equation}
\ddot\theta + 3 \frac{\dot{a}}{a} \dot\theta - \frac{1}{a^2} \nabla^2 \theta =0,
\end{equation}
and the momentum components are $\Pi^{\alpha}\equiv -{3\over 4} \theta^{\alpha}$, so that the relativistic quantum algebra is\cite{RB}
\begin{equation}\label{con}
\left[\theta(x),\theta^{\alpha}(y) \right] =- i \Theta^{\alpha}\, \delta^{(4)} (x-y), \qquad \left[\theta(x),\theta_{\alpha}(y) \right] =
i \Theta_{\alpha}\, \delta^{(4)} (x-y),
\end{equation}
with $\Theta^{\alpha} = i \hbar\, \hat{U}^{\alpha}$ and $\Theta^2 = \Theta_{\alpha}
\Theta^{\alpha} = \hbar^2 \hat{U}_{\alpha}\, \hat{U}^{\alpha}$ for the Riemannian components of velocities $\hat{U}^{\alpha}$.
By making the map $\upsilon = a^{3/2} \theta$, we obtain the dynamic equation for the modes of $\upsilon$
\begin{equation}
\ddot{\upsilon}_k + \omega^2_k(t) \, \upsilon_k =0,
\end{equation}
where the frequency for each mode with wavenumber $k$ is time dependent
\begin{equation}
\omega^2_k(t) = \left[ \frac{k^2}{\beta^2 t^{2p}} - \frac{3 p}{2 t^2} \left( \frac{3}{2} p - 1\right)\right].
\end{equation}
Notice that when $\omega^2_k(t)>0$ the modes are stable, but when $\omega^2_k(t)<0$ they are unstable.
This is the case of the super-Hubble modes during inflation. Since $\theta$ describes a geometrical (quantum) Weylian-like evolution of space-time, the unstable modes will describe the spill of space-time which has been created at quantum scales with wavelength $k$. During this processes the modes suffer a quantum-to-classical transition\cite{prd96}.

The general solution for $\xi_k(t)$ is
\begin{eqnarray}
\xi_k(t) &= &t^{-3 p/2}\, \left\{ A_1 \, {\cal H}^{(1)}_{\nu}\left[y(t)\right]
 +  A_2 \, {\cal H}^{(2)}_{\nu}\left[y(t)\right]\right\},
\end{eqnarray}
where ${\cal H}^{(1,2)}_{\nu}\left[y(t)\right]$ are the Hankel functions, $\nu = { 3p -1\over 2 (p-1)}$, $y(t) = { k \,t^{(1-p)} \over \beta (p-1)}$ and $\beta ={a_0\over t^p_0}$, such that $t_0$ is the time for which inflation starts and
$a_0$ is the initial scale factor of the universe. In order to inflation can take place, we must require that $ p \gg 1$. We must require that the field to be quantized, so
that we obtain the following conditions for the modes $\xi_k(t)$:
\begin{equation}\label{cc}
\xi_k(t) \dot\xi^*_k(t) - \dot\xi_k(t) \xi^*_k(t) = \frac{i}{a^3(t)},
\end{equation}
so that the normalized modes that comply with the condition (\ref{cc}), are
\begin{equation}
\xi_k(t) = \sqrt{\frac{\pi}{4 (p-1)}} t^{-(3p-1)/2} \,{\cal H}^{(2)}_{\nu}[y(t)].
\end{equation}

\subsection{Exact Energy density fluctuations}

In order to calculate the energy density fluctuations during power-law inflation we must calculate $\delta \hat{T}_{\alpha\beta}$ on the Weylian-like manifold. We shall consider that
the variation on the semi Riemannian manifold is null: $\Delta  \hat{T}_{\alpha\beta}=0$, so that
\begin{equation}
\frac{\delta  \hat{T}_{\alpha\beta}}{\delta S} = T_{\alpha\beta|\gamma} U^{\gamma},
\end{equation}
where
\begin{equation}
\hat{T}_{\alpha\beta} = 2 \frac{\delta\hat{\cal L}}{\delta g^{\alpha\beta}} - g_{\alpha\beta} \hat{\cal L},
\end{equation}
and $dx^{\gamma} = U^{\gamma} dS$ is the displacement of any component on the Weylian-like manifold and $\hat{T}_{\alpha\beta}$ are the background (semi Riemannian)
components of the stress tensor. If we take component $\hat{T}_{00} \equiv \hat{\rho}$, we obtain that
\begin{equation}\label{de}
\frac{1}{\hat{\rho}} \frac{\delta \hat{\rho}}{\delta S} = - 2 \theta_0 =  -2 \dot{\theta},
\end{equation}
such that $\dot{\theta} = \left< \dot\theta^2 \right>^{1/2}$.
In order to calculate (\ref{de}), we must find the time derivative of the temporal modes
\begin{equation}
\dot{\xi}_k(t) = \frac{1}{2} \sqrt{\frac{\pi}{p-1}} t^{-2p} \left[ (1-3p) {\cal H}^{(2)}_{\nu}\left[y(t)\right] + \frac{k}{\beta} {\cal H}_{\nu_1}\left[y(t)\right]\right],
\end{equation}
where $\nu_1= \frac{5p-3}{2(p-1)}$. On large (cosmological) scales, the argument of the Hankel functions is very samll: $y(t) \ll 1$, because it takes into account only the
modes with very small wavenumbers. Hence, in order to make an estimation of the spectrum on cosmological scales, will be sufficient with the asymptotic solutions of the Hankel functions:
\begin{equation}
\left.{\cal H}^{(2)}_{\mu}\left[y(t)\right] \right|_{y\ll 1} \simeq \frac{\left[y(t)/2\right]^{\mu}}{\Gamma(1+\mu )} \pm \frac{i}{\pi} \Gamma(\mu) \left[y(t)/2\right]^{-\mu},
\end{equation}
so that
\begin{equation}
\left. \dot{\xi}_k(t) \dot{\xi}^*_k(t) \right|_{y \ll 1}  \simeq  \frac{k^{-2\nu}}{\pi (p-1) (\beta t)^2} \,
\left[ \Gamma(\nu_1) \left[2(p-1) \beta\right]^{\nu_1} + (1-3p) \beta \Gamma(\nu) \, \left[2(p-1) \beta\right]^{\nu} \right]^2.
\end{equation}
The large scales square fluctuations $\left< \dot\theta^2 \right>$, are given by
\begin{equation}
\left.\left< \dot\theta^2 \right>  \right|_{y \ll 1} \simeq  \int^{\epsilon k_0(t)}_0 \, \frac{dk}{k} {\cal P}_{\dot\theta}(k,t),
\end{equation}
where the power-spectrum on cosmological scales is
\begin{equation}
{\cal P}_{\dot\theta}(k,t)  =   \frac{1}{2\pi^2}  \frac{k^{3-2\nu}}{\pi (p-1) (\beta t)^2}  \,
\left[ \Gamma(\nu_1) \left[2(p-1) \beta\right]^{\nu_1} + (1-3p) \beta \Gamma(\nu) \, \left[2(p-1) \beta\right]^{\nu} \right]^2.
\end{equation}
The spectral index is given by $n_s-1=3-2\nu$, so that once known $n_s$, it is possible to obtain the power of the expansion of the universe
\begin{equation}
p= 1+\frac{2}{1-n_s}.
\end{equation}
For $n_s \simeq 0.96$\cite{RPP1}, one obtains $p \simeq 51$. For this spectral index, the equation of state  $P = -\left(\frac{2\dot H}{3 H^2} +1 \right)\,\rho$, results to be
\begin{equation}
\omega= \frac{2-3p}{3p} \simeq -0.9869,
\end{equation}
which agrees with WMAP9-$\omega$CDM(flat) observations\cite{RPP2}.
where $P$ is the pressure and $\rho$ the energy density. Furthermore, this value corresponds to a ratio for the tensor to scalar indices
\begin{equation}
r \equiv \frac{\Delta^2_t(k_*)}{\Delta^2_s(k_*)}\simeq 0.106.
\end{equation}
This value is in agreement with that one expects for inflationary models.

\section{Final remarks}

In Relativistic Quantum Geometry, the dynamics of a geometrical scalar field defined in a Weylian-like integrable
manifold preserves the gauge-invariance under transformations of the Einstein equations and the geometrical vector fields, that involves the cosmological constant. The scalar field $\theta$, always can be quantized because it is free of interactions, and describes a relativistic quantum algebra: $\left[\theta(x), \theta^{\mu}(y)\right] =-i
\, \Theta^{\mu} \delta^{(4)}(x-y)$. In this letter we have examined an example in the framework of a power-law inflationary cosmology. The results agree perfectly with observations.
An interesting prospect should be the calculation of vector geometrical field dynamics during inflation:
\begin{equation}\label{mx}
\hat\Box A^{\nu} -  \hat\nabla^{\nu} \left(\hat\nabla_{\mu} A^{\mu} \right) = J^{\nu},
\end{equation}
which would describe the seed of electromagnetic fields under the influence geometrical currents, $J^{\nu}$, produced by geometrical charged fields. For a co-moving relativistic observer
the gauge equations must be $\hat\nabla_{\mu} A^{\mu} =\frac{\Lambda}{2} \theta_{\mu} \hat{U}^{\mu}$, with $U^{\mu} =(1,0,0,0)$. However, these calculations are beyond the scope of this letter.

\section*{Acknowledgements}

\noindent  M. Bellini acknowledges
UNMdP and  CONICET (Argentina) for financial support.


\begin{thebibliography}{99}
\bibitem{infl}
A. A. Starobinsky, Phys. Lett. {\bf B91}: 99 (1980).
A. H. Guth, Phys. Rev. {\bf D23}: 347 (1981);
A. D. Linde, Phys. Lett. {\bf B129}: 177 (1983).
\bibitem{ab} M. Anabitarte, M. Bellini, Eur. Phys. J. {\bf C60}: 297-301 (2009).
\bibitem{mio} M. Bellini, Phys. Rev. {\bf D61}: 107301 (2000).
\bibitem{muk} R. Isaacson, Phys. Rev.: 166 (1964); \\
V. F. Mukhanov, L. R. W. Abramo and R. H. Branderberger, Phys. Rev. Lett. {\bf 78}: 1624 (1997).
\bibitem{RB} L. S. Ridao, M. Bellini, {\em Towards relativistic quantum geometry}. E-print: arXiv: 1506.09141; \\
M. R. A. Arcod\'{\i}a, M. Bellini, {\em Charged and electromagnetic fields from relativistic quantum geometry}. E-print: arXiv: 1608.07899.
\bibitem{Weyl} H. Weyl, {\em Philosophy of Mathematics and Natural Science}, english version, Princeton University Press (1949).
\bibitem{rb} L. S. Ridao, M. Bellini, Astrophys. Space Sci., {\bf 357} (2015) 1, 94.
\bibitem{prd96} M. Bellini, H. Casini, R. Montemayor, P. Sisterna, Phys. Rev. {\bf D54}: 7172 (1996).
\bibitem{RPP1} O. Lahov, A. R. Liddle, Chin. Phys. {\bf C38}: 345-352 (2014).
\bibitem{RPP2} M. J. Mortonson, D. H. Weinbert, M. White, Chin. Phys. {\bf C38}: 361-368 (2014).
\end{thebibliography}
\end{document}